\documentclass[twocolumn]{aastex631}

\usepackage{amsmath}	
\usepackage{amssymb}	

\DeclareRobustCommand{\ion}[2]{%
\relax\ifmmode
\ifx\testbx\f@series
{\mathbf{#1\,\mathsc{#2}}}\else
{\mathrm{#1\,\mathsc{#2}}}\fi
\else\textup{#1\,{\mdseries\textsc{#2}}}%
\fi}

\usepackage{graphicx}	
\usepackage{subfigure}

\received{February 22, 2023}
\revised{April 12, 2023}
\accepted{April 20, 2023}

\submitjournal{ApJL}

\shorttitle{CSS Sources with Enhanced SF are Smaller than $10\,$kpc}
\shortauthors{Y.~A.~Gordon et al.}
\graphicspath{{./}{Figures/}}

\begin{document}

\title{Compact Steep Spectrum Radio Sources with Enhanced Star Formation are Smaller than $10\,$kpc}

\correspondingauthor{Yjan~A. Gordon}
\email{yjan.gordon@wisc.edu}

\author[0000-0003-1432-253X]{Yjan~A. Gordon}
\affil{Department of Physics, University of Wisconsin-Madison, 
1150 University Ave, Madison, WI 53706, USA}

\author[0000-0001-6421-054X]{Christopher~P. O'Dea}
\affiliation{Department of Physics and Astronomy, University of Manitoba, 
Winnipeg, MB R3T 2N2, Canada}

\author[0000-0002-4735-8224]{Stefi~A. Baum}
\affil{Department of Physics and Astronomy, University of Manitoba, 
Winnipeg, MB R3T 2N2, Canada}

\author[0000-0001-8156-0429]{Keith Bechtol}
\affil{Department of Physics, University of Wisconsin-Madison, 
1150 University Ave, Madison, WI 53706, USA}

\author{Chetna Duggal}
\affiliation{Department of Physics and Astronomy, University of Manitoba, 
Winnipeg, MB R3T 2N2, Canada}

\author[0000-0001-6957-1627]{Peter~S. Ferguson}
\affil{Department of Physics, University of Wisconsin-Madison, 
1150 University Ave, Madison, WI 53706, USA}




\begin{abstract}
Compact Steep Spectrum (CSS) radio sources are active galactic nuclei that have radio jets propagating only on galactic scales, defined as having projected linear sizes (LS) of up to $20\,$kpc.
CSS sources are generally hosted by massive early-type galaxies with little on-going star formation, however a small fraction are known to have enhanced star formation.
Using archival data from the Faint Images of the Radio Sky at Twenty cm survey, the Very Large Array Sky Survey and the Sloan Digital Sky Survey we identify a volume-limited sample of $166$ CSS sources at $z<0.2$ with $L_{1.4\,\text{GHz}}>10^{24}\,\text{W}\,\text{Hz}^{-1}$.
Comparing the star formation rates and linear sizes of these CSS sources, we find that the $\approx14\,\%$ of CSS sources with specific star formation rates above $0.01\,\text{Gyr}^{-1}$ all have $\text{LS}<10\,$kpc.
We discuss the possible mechanisms driving this result, concluding that it is likely the excess star formation in these sources occurred in multiple bursts and ceased prior to the AGN jet being triggered.
\end{abstract}

\keywords{
\href{http://astrothesaurus.org/uat/16}{Active galactic nuclei (16)},
\href{http://astrothesaurus.org/uat/2017}{AGN host galaxies (2017)},
\href{http://astrothesaurus.org/uat/508}{Extragalactic radio sources (508)},
\href{http://astrothesaurus.org/uat/1343}{Radio Galaxies (1343)},
\href{http://astrothesaurus.org/uat/1569}{Star formation (1569)},}



\section{Introduction} 
\label{sec:intro}

Active Galactic Nuclei (AGN) are the phenomenon whereby matter is accreting onto the central supermassive black hole of their host galaxies \citep{Salpeter1964}.
A small fraction of AGN produce particle jets that result in radio emission via mechanisms such as synchrotron radiation and inverse Compton scattering
\citep{Padovani2017, Blandford2019}.
The jets produced by these radio loud AGN (RLAGN) can sometimes propagate well beyond the host galaxy, giving rise to large scale double-lobed structures such as Fanaroff and Riley class I and II radio galaxies \citep[FRIs and FRIIs,][]{Fanaroff1974} that can span hundreds of kiloparsecs or more \citep[e.g.,][]{Willis1974, IshwaraChandra1999, Dabhade2017}.
In contrast to FRIs and FRIIs are compact RLAGN that have radio emission on scales similar to or smaller than the host galaxy. 

Compact Steep Spectrum (CSS) radio sources have radio extents smaller than $\sim20\,$kpc and radio spectral indices of $\alpha < -0.5$, where spectral index, $\alpha$, is related to flux density, $S$, and frequency, $\nu$, by $S \propto \nu^{\alpha}$
\citep{Fanti1990, ODea1998, ODea2021}.
It is thought that at least some CSS sources are young AGN that will evolve into larger radio morphologies \citep{Fanti1995, ODea1998, An2012, ODea2021}.
This hypothesis is based on very long baseline interferometry (VLBI) observations of powerful CSS sources that show double-lobed radio morphologies analogous to FRIs and FRIIs but on a much smaller scale \citep{Spencer1991, Dallacasa1995}, and jet proper motions indicative of a short travel time from the central engine \citep{Owsianik1998, Polatidis2003, An2012propermotions}.
The young AGN scenario is further supported by CSS sources having host galaxies similar those of larger radio galaxies.

An alternative to the young AGN scenario, is that the radio jets in CSS sources are unable to travel as easily through the interstellar medium  (ISM), a phenomenon known as `frustration' \citep{vanBreugel1984, Wilkinson1984, ODea1991}.
Frustration can occur either as a result of intrinsically weak jet power, the jet strongly interacting with a dense ISM, or a combination of these factors.
The jet frustration paradigm is supported by high resolution images that show distinct asymmetry in some CSS radio sources \citep{Saikia1995, Saikia2003, Orienti2007}.

The galaxies that host CSS sources are generally massive early-type galaxies with little star formation (SF), but a subset of the CSS population are known to exhibit enhanced SF \citep{deVries1998, deVries2000, Drake2004, Tadhunter2011, Dicken2012, ODea2021}.
A systematic study of SF in a large sample of CSS host galaxies may help shed light on to why some CSS sources are star forming and constrain the evolutionary path of these RLAGN.
Such studies have previously been problematic as most samples of CSS sources consisted of objects with high radio luminosity that are rare in the local Universe.
Consequently, the relatively shallow wide-field multiwavelength surveys that can readily provide star formation rates (SFRs) usually don't cover the high luminosity radio sources and expensive targeted observations are often necessary.
The advent of deep, wide-field radio continuum surveys with high angular resolution is now making these types of systematic studies feasible \citep{Sadler2016}.

The Faint Images of the Radio Sky at Twenty cm survey \citep[FIRST,][]{Becker1995} and the Very Large Array Sky Survey \citep[VLASS,][]{Lacy2020}, which have angular resolutions of $5.4''$ and $3''$ respectively, are well suited to identifying compact radio sources brighter than $\approx 1\,$mJy.
Furthermore, these surveys observe at different frequencies; FIRST at $1.4\,$GHz and VLASS at $3\,$GHz.
Using FIRST and VLASS data together is a pragmatic approach to measuring the spectral indices of large numbers of faint compact radio sources \citep{Gordon2021}.
Both FIRST and VLASS cover the $\approx 10,000\,\text{deg}^{2}$ footprint of the Sloan Digital Sky Survey \citep[SDSS,][]{York2000} which provides optical measurements and derived properties, including SFRs, for $\sim 10^{6}$ galaxies.
Combining FIRST, VLASS and SDSS therefore has the potential to be an effective method for studying SF in a large number of CSS sources in the local Universe.

In this Letter we use data from FIRST, VLASS and SDSS to investigate the relationship between radio source size and SF in CSS sources.
The selection of CSS sources is described in Section \ref{sec:data}.
In Section \ref{sec:results} we compare the radio sizes and SFRs of our CSS sources.
We discuss our results in Section \ref{sec:discussion} and state our conclusions in Section \ref{sec:conclusions}.
Throughout this work we assume a flat $\Lambda$CDM cosmology with $h=0.7$, $H_{0} = 100h\,\text{km}\,\text{s}^{-1}\,\text{Mpc}^{-1}$, $\Omega_{m}=0.3$ and $\Omega_{\Lambda}=0.7$.


\section{Sample Selection} 
\label{sec:data}

To identify likely CSS sources we start with the \citet{Best2012}
catalog of radio galaxies in the SDSS 7th Data Release \citep[DR7,][]{Abazajian2009} spectroscopic sample.
This catalog contains the host IDs of the radio sources, identifies sources where the radio emission is likely due to SF rather than an AGN, and where possible classifies RLAGN as either low- or high-excitation radio galaxies (LERGs or HERGs).
As we are interested in compact radio AGN in this work, we select objects from the \citet{Best2012} catalog that are associated with a single detection in FIRST, excluding multi-FIRST-component sources from consideration.

High frequency ($\nu\sim3\,$GHz) information on our sources is obtained by cross matching with the VLASS Epoch 1 component catalog \citep{Gordon2021}.
We only search for VLASS components brighter than $3\,$mJy/beam, as fainter components have less reliable flux density measurements \citep[See section 3 of][]{Gordon2021}.
A search radius of $5''$ is used which, given the on-sky component density of VLASS at $S > 3\,$mJy ($\sim18\,\text{deg}^{-2}$), has an expected contamination level from false-positive matches of less than $0.05\,\%$.
The $3\,$GHz flux density of our sources is then scaled by $1/0.87$ to account for the systematic underestimation of flux density measurements in the VLASS catalog reported in \citet{Gordon2021}.
With flux densities at two different frequencies in hand, we determine the spectral index, $\alpha$, between $1.4\,$GHz and $3\,$GHz for our sources. 

The projected radio extents of CSS sources are smaller than $20\,$kpc \citep[e.g.][]{Fanti1985, ODea1997, ODea2021}.
The VLASS catalog of \citet{Gordon2021} includes measurements of the source angular size after after deconvolution from the beam\footnote{These measurements are produced by the source-finder PyBDSF \citep{Mohan2015}.}.
Where the deconvolved angular size is non-zero, this is used to calculate the projected linear size (LS) of the source.
If the source is so compact that it has a deconvolved angular size of zero in VLASS, then we use the uncertainty in the angular size to estimate an upper limit on the LS.

\begin{figure}
    \centering
    \includegraphics[width=\columnwidth]{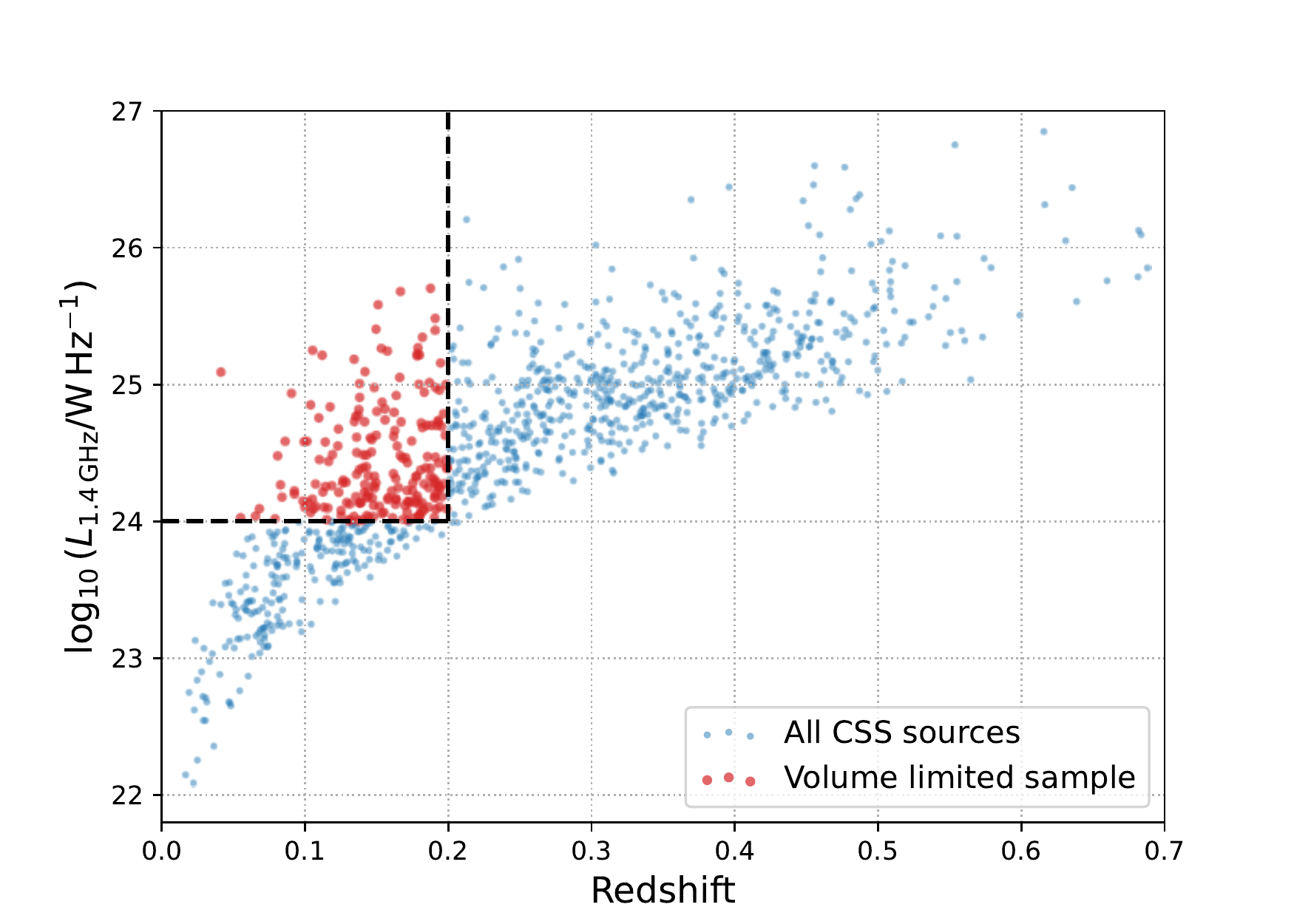}
    \caption{Redshift and $1.4\,$GHz luminosity distributions for our selection of CSS sources.
    Our volume limited sample (red circles) is defined as having $z<0.2$ and \text{$L_{1.4\,\text{GHZ}}>10^{24}\,\text{W}\,\text{Hz}^{-1}$.}
    }
    \label{fig:z_v_lum}
\end{figure}

We select our likely CSS sources as having $\text{LS} < 20\,$kpc and $\alpha + \sigma_{\alpha} < -0.5$, identifying $1,109$ objects.
In \text{Figure \ref{fig:z_v_lum}} we compare the redshifts and $1.4\,$GHz luminosities of this sample.
By using only sources at $z<0.2$ we select a volume-limited sample complete down radio luminosities of $L_{1.4\,\text{GHZ}} > 10 ^{24}\,\text{W}\,\text{Hz}^{-1}$.
This sample contains $259$ CSS candidates, all but $38$ ($15\,\%$) of which are classified as LERGs.
The radiatively efficient central engines in HERGs can impact the observed properties of the host galaxy, including spectral line measurements used in determining SFRs.
Conversely, the radiatively inefficient central engines of LERGs don't produce the high energy photons necessary to bias spectral line measurements \citep{Hardcastle2006}. 
We therefore exclude the $38$ sources not classified as LERGs.
Finally, $\approx 20\,\%$ of single-FIRST-component RLAGN are expected to be multi-component sources in VLASS \citep{Gordon2019}.
To ensure we are only using sources with reliable sizes and spectral indices, we visually inspect the VLASS maps using SAOImage DS9 \citep{Joye2003} with the catalog components overlaid. 
As a result we remove $55$ multi-VLASS-component sources from our sample, leaving $166$ CSS sources that we use for the analysis presented in this Letter.

\section{Comparing Star Formation and Linear Size in CSS Sources} 
\label{sec:results}

\begin{figure}
    \centering
    \includegraphics[width=\columnwidth]{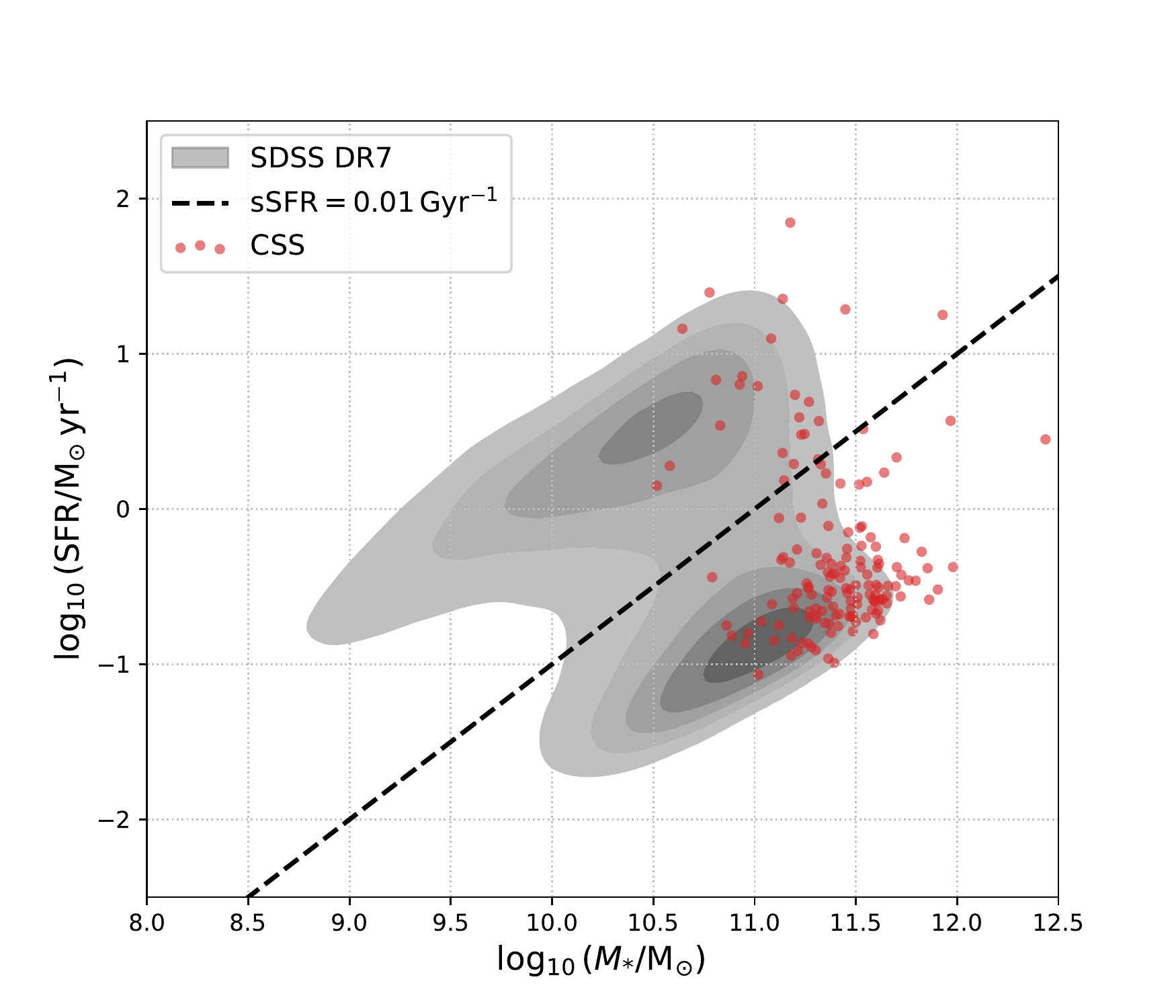}
    \caption{The stellar masses and star formation rates of our CSS sources (red circles).
    The grey shaded contours show the distributions for SDSS DR7.
    The black dashed line shows a fixed specific star formation rate of $0.01\,\text{Gyr}^{-1}$.
    }
    \label{fig:mass_sfr}
\end{figure}

All of our CSS sources have host galaxies with spectral line measurements, stellar masses ($M_{*}$) and SFRs in the Max Plank institut f\"ur Astrophysik/Johns Hopkins University (MPA/JHU) value added catalog for SDSS DR7\footnote{\url{https://wwwmpa.mpa-garching.mpg.de/SDSS/DR7/}} \citep{Kauffmann2003, Brinchmann2004, Tremonti2004}.
In Figure \ref{fig:mass_sfr} we plot the stellar masses and SFRs of our CSS sources.
For reference we also show the distribution of all galaxies at $z<0.2$ in SDSS DR7 as grey shaded contours.
Galaxies in SDSS are split into two populations of `star-forming' and `passive' at a specific star formation rate ($\text{sSFR} = \text{SFR}/M_{*}$) of approximately $0.01\,\text{Gyr}^{-1}$ (shown by the black dashed line in Figure \ref{fig:mass_sfr}).
The majority of our CSS sources are hosted by passive high-mass galaxies, with only $24\ (14\,\%)$ having $\text{sSFR} > 0.01\,\text{Gyr}^{-1}$.
We confirm these have a similar redshift distribution to the passive CSS hosts in our sample by performing a Kolmogorov-Smirnov (KS) test, which returns a $p-$value of $0.75$.

\begin{figure*}
    \centering
    \subfigure[]{\includegraphics[trim={4mm, 0, 4mm, 0}, clip, width=\columnwidth]{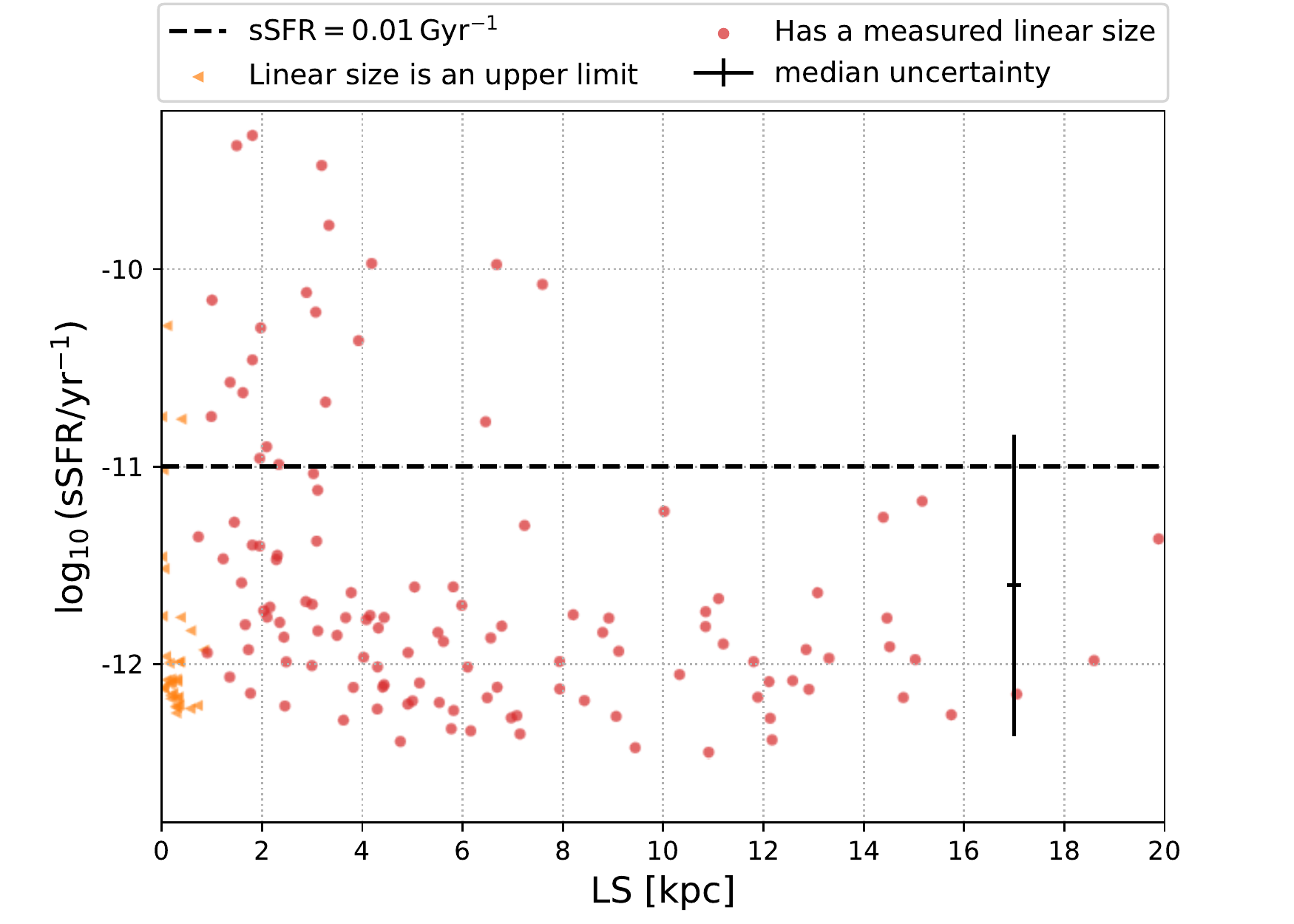}}
    \hspace{5mm}
    \subfigure[]{\includegraphics[trim={4mm, 0, 4mm, 0}, clip, width=\columnwidth]{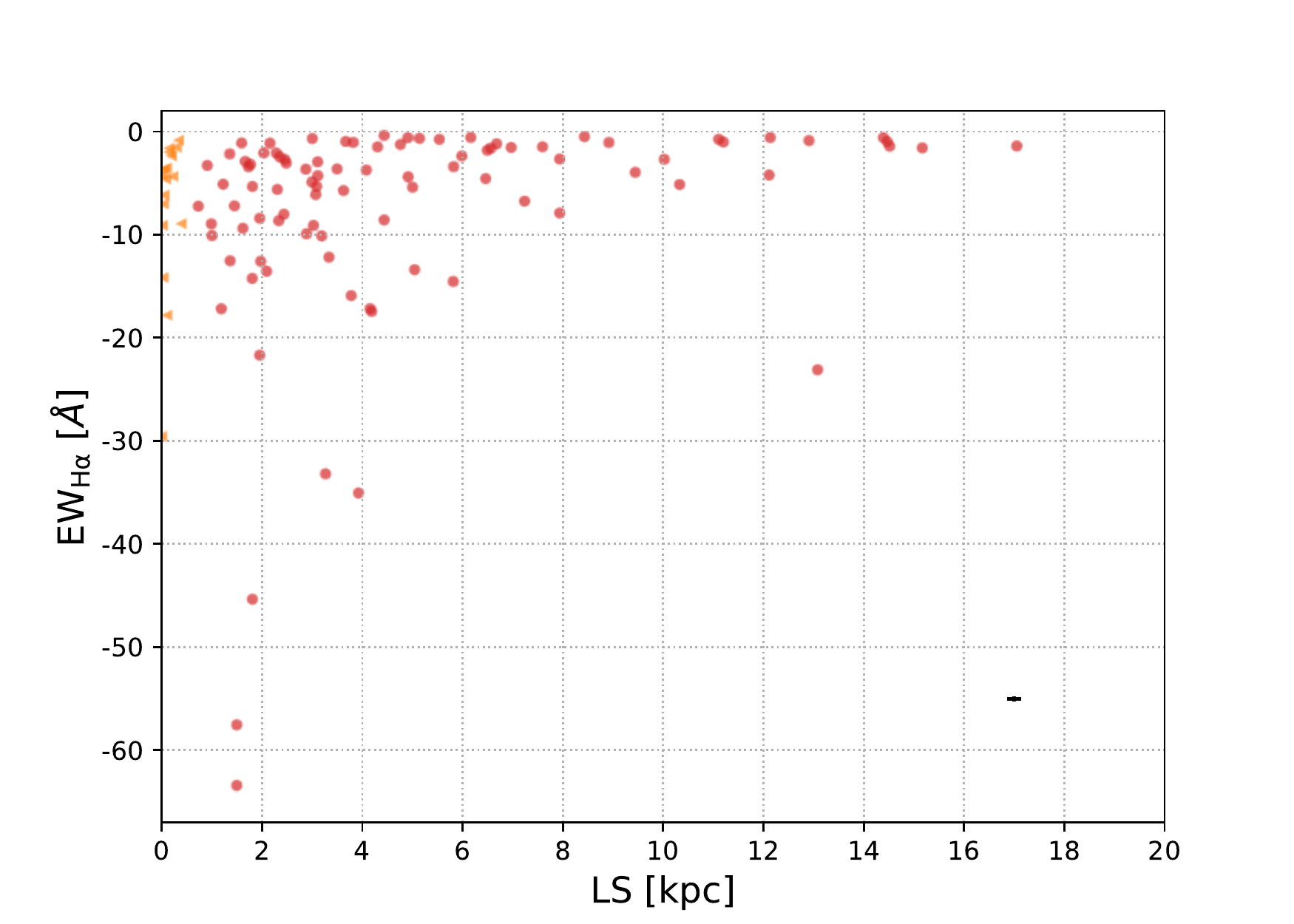}}
    \subfigure[]{\includegraphics[trim={4mm, 0, 4mm, 0}, clip, width=\columnwidth]{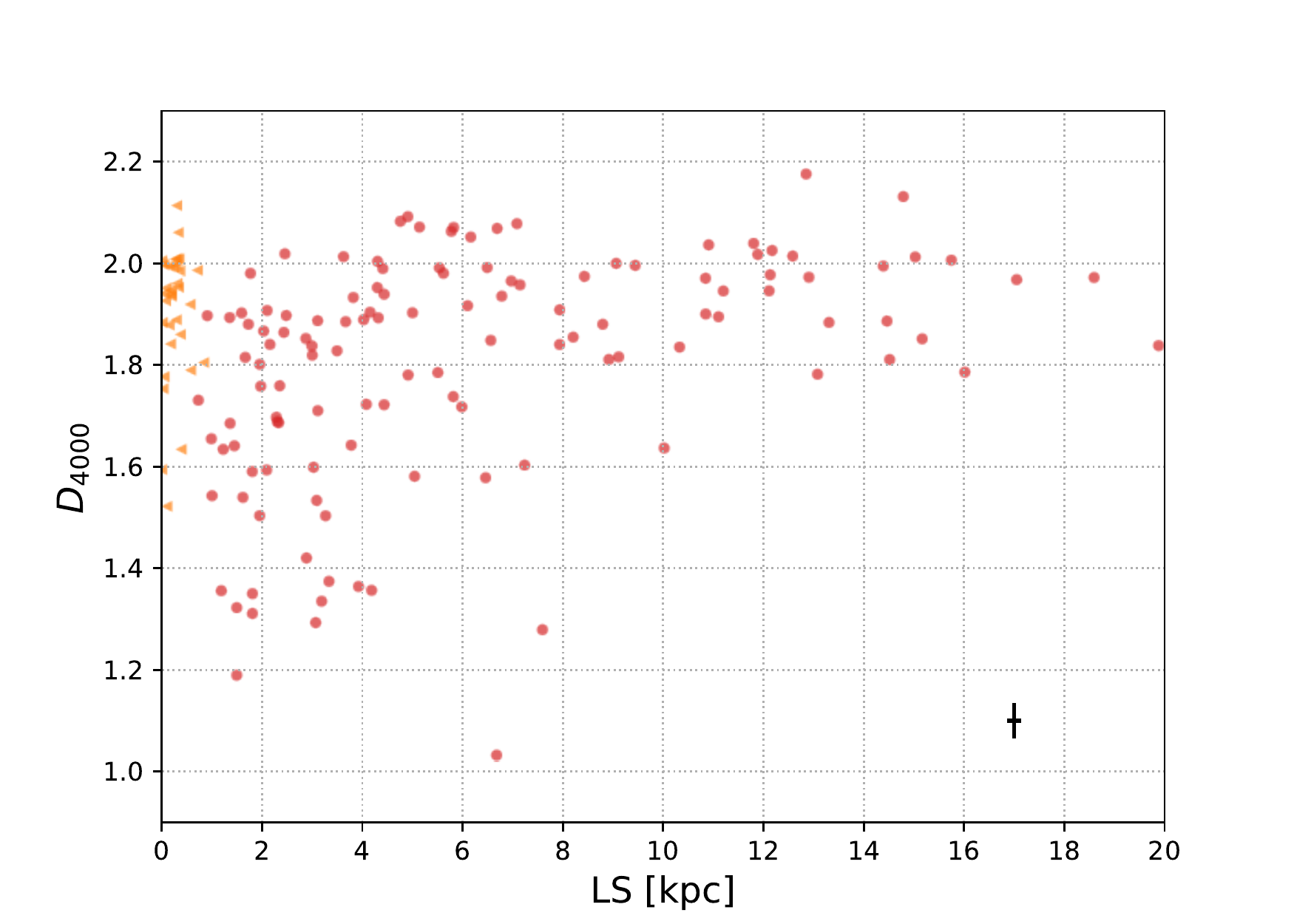}}
    \hspace{5mm}
    \subfigure[]{\includegraphics[trim={4mm, 0, 4mm, 0}, clip, width=\columnwidth]{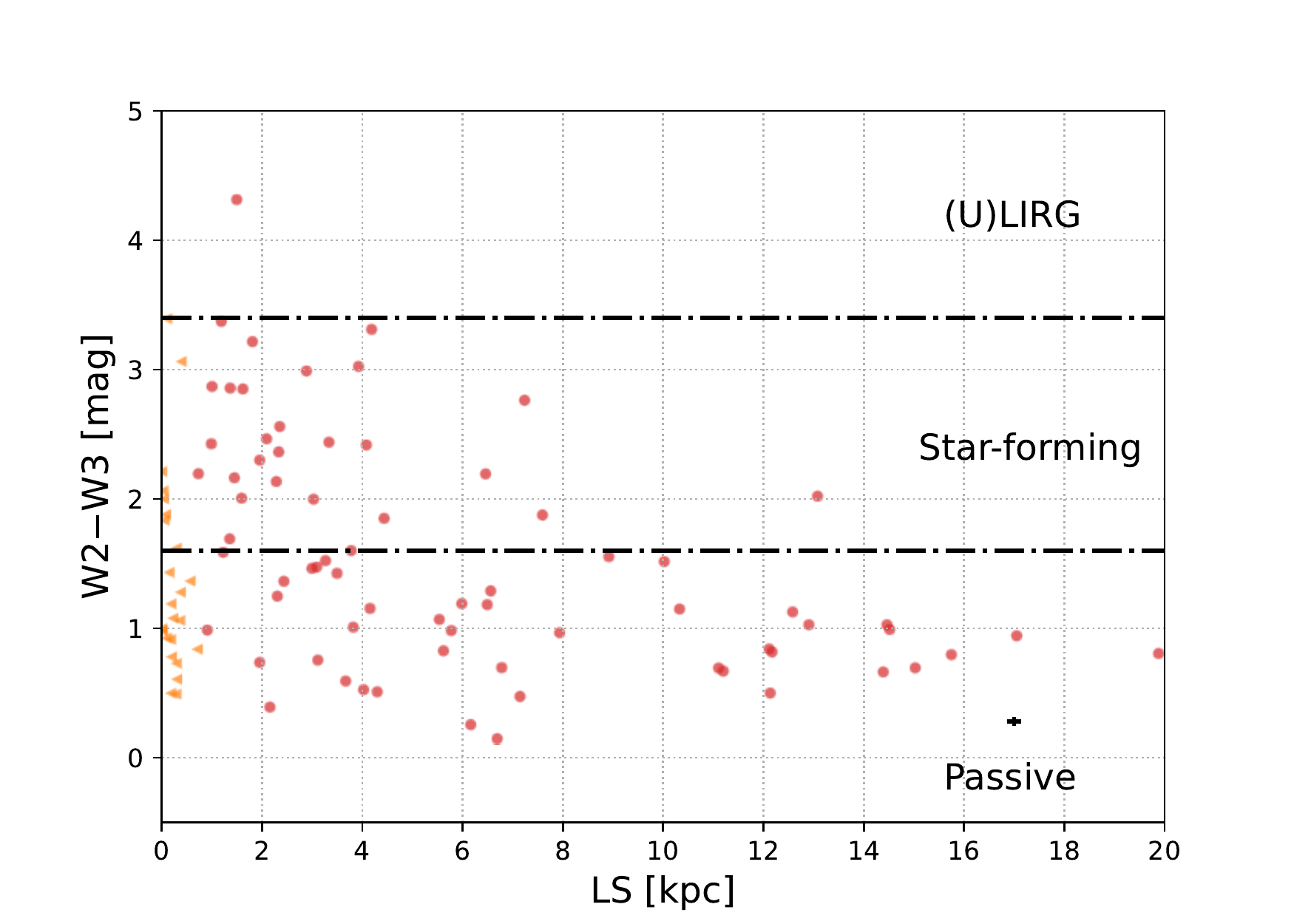}}
    \caption{
    Comparisons of star-forming indicators and linear size (LS) for our CSS sources.
    Panel a shows the SDSS sSFR measurements for our CSS sources with a black dashed line indicating $\text{sSFR}=0.01\,\text{Gyr}^{-1}$.
    Panel b shows the equivalent width of H$\alpha$ ($\text{EW}_{\text{H}\alpha}$) for our CSS sources where this line is detected at $S/N > 3$, while panel c shows the strength of the $4,000\,$\AA\ break.
    In Panel d the WISE colors are shown for galaxies where the AGN does not dominate the IR color ($\text{W1}-\text{W2} < 0.5$).
    Here the dot-dashed lines separate colors associated with passive galaxies, star-forming galaxies and (U)LIRGs.
    In all panels orange triangles denote CSS sources where the LS is an upper limit, and the black cross shows the median uncertainty for the data points.}
    \label{fig:size_v_ssfr}
\end{figure*}

With SF being rare in CSS sources, one might ask whether there are differences between CSS sources with SF and CSS sources hosted by passive galaxies?
One of the most fundamental properties of RLAGN is their size, i.e., how far the jets have travelled from the central engine.
To assess if the sizes of CSS sources with SF and passive hosts differ we plot the LS of our CSS sources versus their host sSFR in Figure \ref{fig:size_v_ssfr}a.
CSS sources with $\text{sSFR}<0.01\,\text{Gyr}^{-1}$ are seen at all sizes in our sample (\text{$0 < \text{LS} < 20\,$kpc}).
However, higher \text{sSFRs} ($\text{sSFR}>0.01\,\text{Gyr}^{-1}$) are only seen in CSS sources with  $\text{LS}\lesssim 8\,$kpc.
If we divide the CSS population at $\text{LS}=10\,$kpc, $17.9_{-2.8}^{+3.8}\,\%$ of sources with $\text{LS}<10\,$kpc have $\text{sSFR}>0.01\,\text{Gyr}^{-1}$, compared to $0.0_{-0.0}^{+0.5}\,\%$ at $\text{LS}\geq10\,$kpc.
The uncertainties in these population fractions are estimated using the binomial approach outlined in \citet{Cameron2011}, and suggest a $\approx 2.9\sigma$ excess of star-forming hosts in the smaller CSS sources.

The SFR measurements in SDSS are based on either the H$\alpha$ luminosity or the strength of the $4,000\,\text{\AA}$ break, $D_{4000}$, depending on the spectral line properties of the galaxy \citep{Brinchmann2004}.
In panels b and c of Figure \ref{fig:size_v_ssfr} we show both of these observable properties complement our findings with respect to the derived sSFRs shown in Figure \ref{fig:size_v_ssfr}a.
Where H$\alpha$ is detected ($S/N > 3$), the strongest H$\alpha$ emission lines are found nearly exclusively in CSS sources with $\text{LS}<6\,$kpc (Figure \ref{fig:size_v_ssfr}b).
When considering $D_{4000}$, the weakest breaks--indicating young stellar populations--are found only in CSS sources with $\text{LS}\lesssim 10\,$kpc (Figure \ref{fig:size_v_ssfr}c).

A further test of the relative compactness of CSS sources with enhanced SF is to investigate how the infrared (IR) colors of the host change with LS.
To this end we obtain IR information from the Wide-field Infrared Survey Explorer telescope \citep[WISE,][]{Wright2010} AllWISE catalog \citep{Cutri2012, Cutri2013}.
The WISE W2 ($4.3\mu$m) and W3 ($12\mu$m) filters can be used to identify star-forming galaxies.
Additionally, the W1 ($3.4\mu$m) and W2 filters can identify galaxies where the IR colors are contaminated by AGN emission. 
From our sample of CSS sources, $102$ ($61\,$\%) are detected ($S/N >2$) in the W1, W2 and W3 bands.
Of these $102$ galaxies, $6$ have $\text{W1}-\text{W2} > 0.5$ indicating that their IR colors are dominated by the AGN \citep{Mingo2016}.
For the remaining $96$ CSS sources, we plot their $\text{W2}-\text{W3}$ color against LS in Figure \ref{fig:size_v_ssfr}d.
Adopting the criteria of \citet{Mingo2016}, galaxies with
\begin{itemize}
    \item $\text{W2}-\text{W3} < 1.6 $ are passive,
    \item $1.6 < \text{W2}-\text{W3} < 3.4$ are star-forming,
    \item and $\text{W2}-\text{W3} > 3.4$ are (Ultra) Luminous Infrared Galaxies ([U]LIRGs).
\end{itemize}
Panel d of Figure \ref{fig:size_v_ssfr} is consistent with panels a-c, showing that nearly all CSS sources with IR colors indicative of SF have $\text{LS}<8\,$kpc. 
For CSS sources with $\text{LS}<10\,$kpc, $40.0_{-5.0}^{+5.5}\,\%$ have star-forming WISE colors.
On the other hand, only $5.9_{-1.9}^{+11.3}\,\%$ of CSS sources with $\text{LS} \geq 10\,$kpc have WISE colors associated with star-forming galaxies--a deficit relative to the sub $10\,$kpc population at $\approx2.8\sigma$ confidence.

\section{Discussion} 
\label{sec:discussion}

\subsection{Physical Interpretation}

Our data shows that where excess SF is present in CSS sources, those sources are limited to scales smaller than $\approx 10\,$kpc.
At first glance, there are three likely possibilities that might explain this observation.
\begin{enumerate}
    \item The jet itself has triggered a brief period of SF \citep[e.g.][]{Rees1989, Labiano2008, Duggal2021}.
    \item A dense ISM is inhibiting the propagation of the radio jet resulting in its confinement to scales \text{$\lesssim 10\,$kpc.}
    \item The AGN is younger than the SF, limiting the time available for the radio jets to propagate away from the central engine. 
\end{enumerate}

\begin{figure}
    \centering
    \subfigure[]{\includegraphics[trim={5mm, 0, 5mm, 0}, clip, width=\columnwidth]{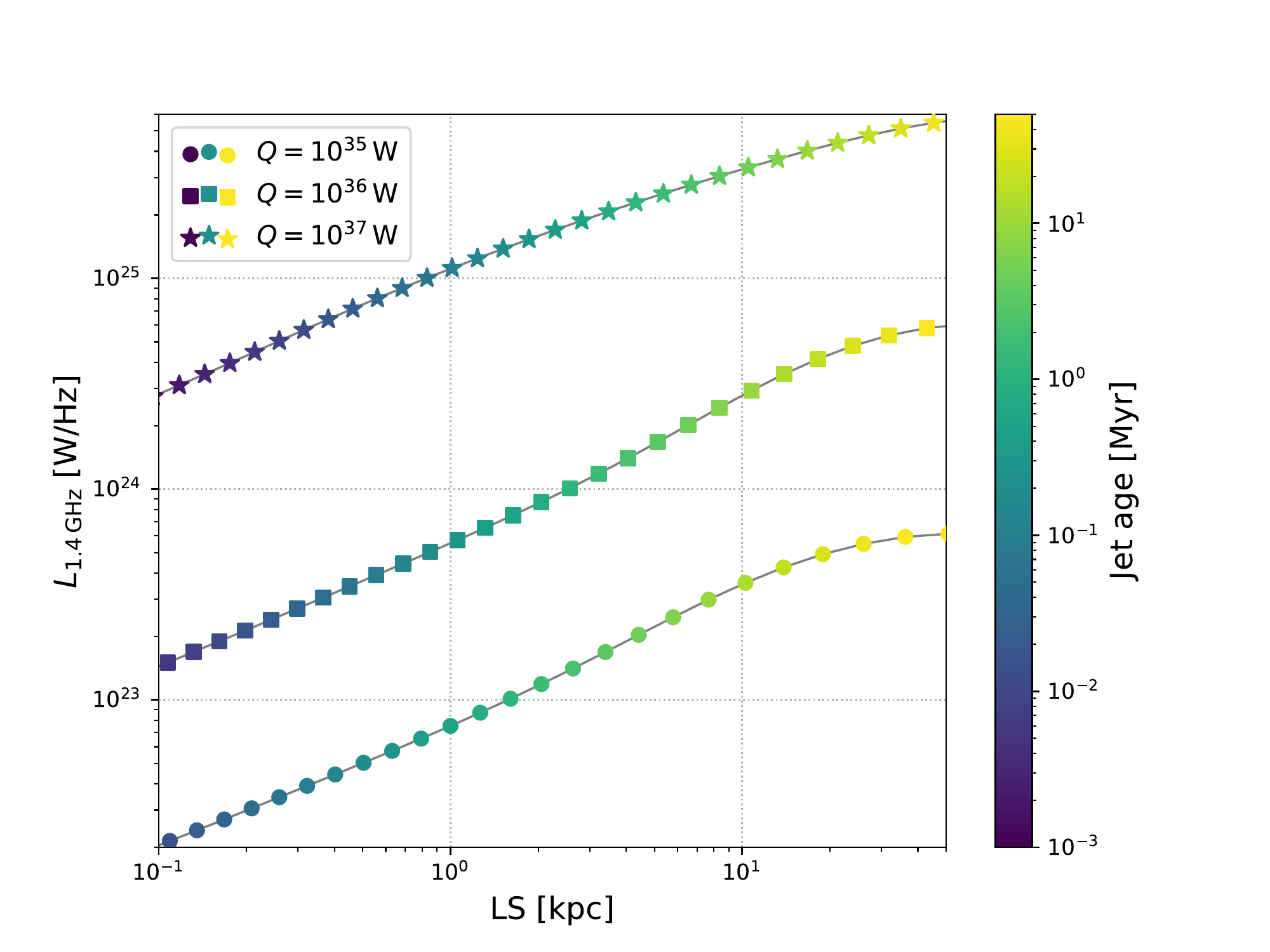}}
    \hspace{5mm}
    \subfigure[]{\includegraphics[trim={5mm, 0, 5mm, 0}, clip, width=\columnwidth]{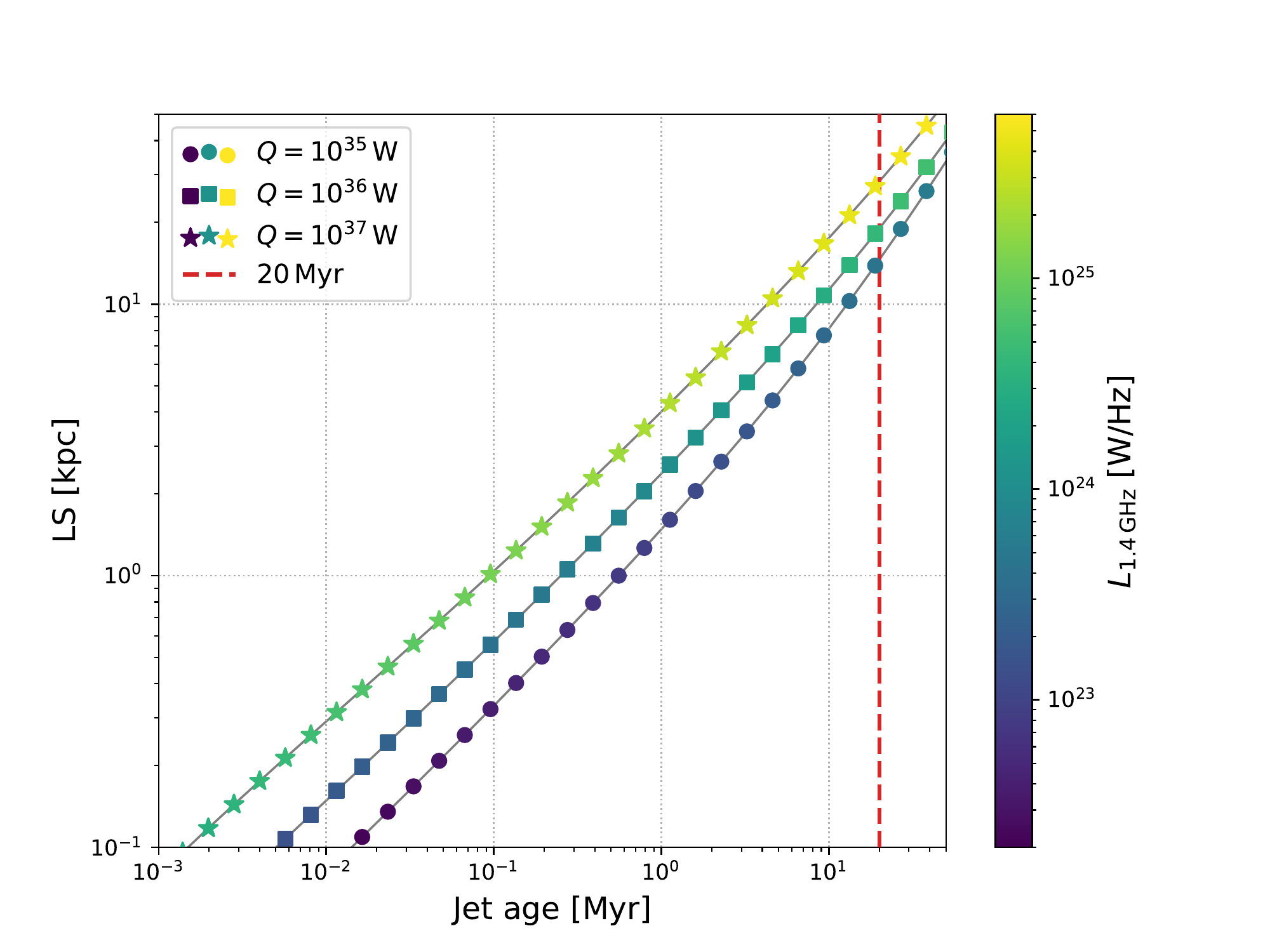}}
    \caption{Three toy model jets simulated using the semi-analytical code of \citet{Hardcastle2018} with powers $Q=10^{35}\,$W (circles), $Q=10^{36}\,$W (squares) and $Q=10^{37}\,$W (stars).
    Panel a shows the evolution of $1.4\,$GHz luminosity with linear size colored by the jet age.
    Panel b shows the linear size growth as a function jet age, with the points colored by $L_{1.4\,\text{GHz}}$.
    The red dashed line in panel b shows the $20\,$Myr typical lifetime of O-type stars.}
    \label{fig:jet_models}
\end{figure}

To explore these scenarios we compare the expected evolution of radio jets in these sources to the timescale on which the SF is detectable.
In order to estimate the typical age of the jets in our CSS sources with enhanced SF, we
simulate three `toy model' jets using the semi-analytical radio jet evolution code of \citet{Hardcastle2018}.
The median $1.4\,$GHz luminosity of our sample is $10^{24.3}\,\text{W}\,\text{Hz}^{-1}$.
For this simulation we assume a universal pressure profile \citep{Arnaud2010} for galaxies in a halo of mass $M_{500}=10^{13.5}\,\text{M}_{\odot}$.
In this scenario, a radio source with $\text{LS}=10\,$kpc and $L_{1.4\,\text{GHz}}=10^{24.3}\,\text{W}\,\text{Hz}^{-1}$ is expected to have a jet power, $Q$, of $\sim 10^{36}\,\text{W}$ (see Figure \ref{fig:jet_models}a).
Such a jet will have taken $\approx8\,$Myr to reach its current size, and would reach a linear size of $\approx19\,$kpc within $20\,$Myr of being switched on (see Figure \ref{fig:jet_models}b).

\begin{figure}
    \centering
    \subfigure[]{\includegraphics[trim={5mm, 0, 5mm, 0}, clip, width=\columnwidth]{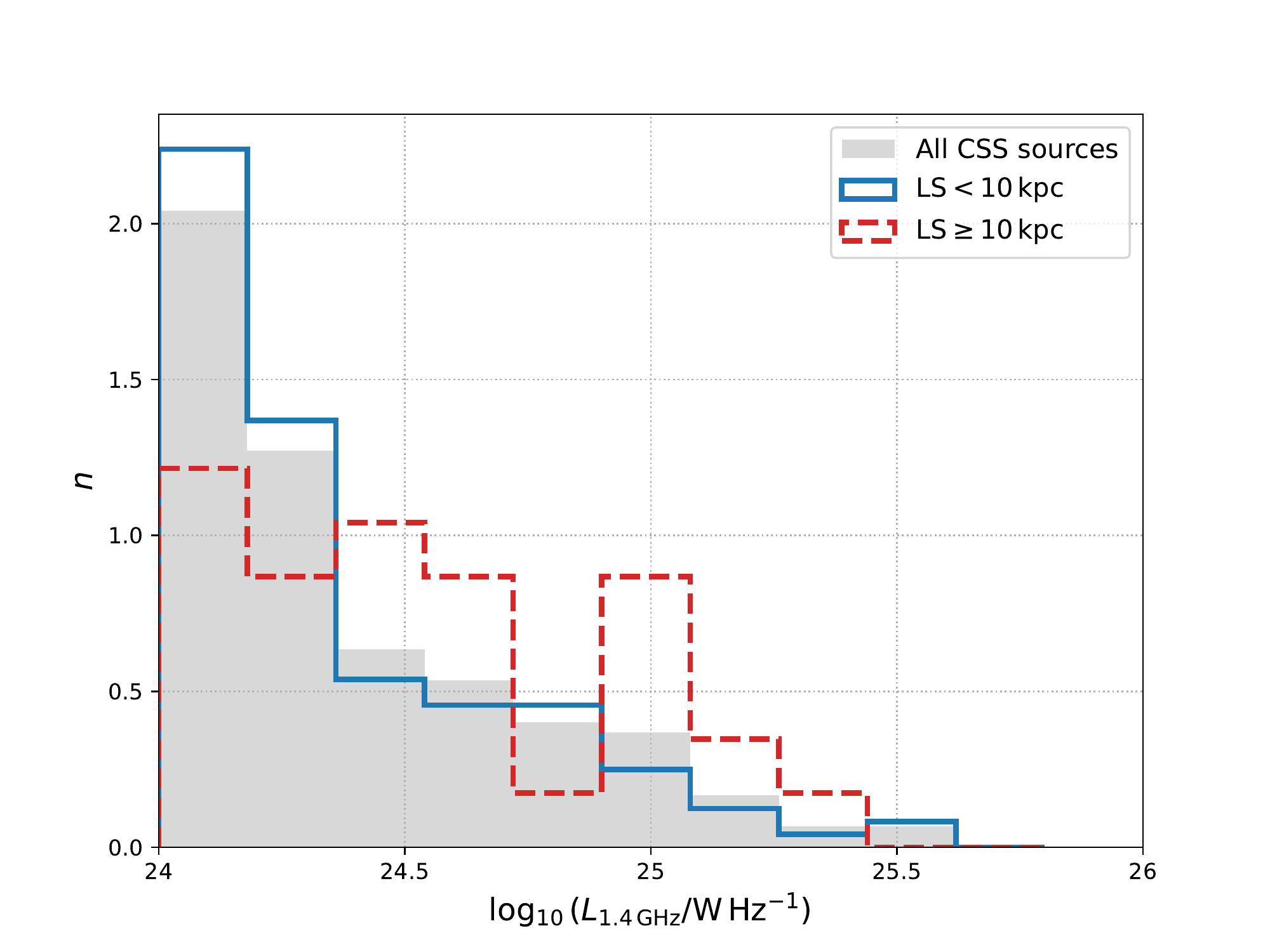}}
    \subfigure[]{\includegraphics[trim={5mm, 0, 5mm, 0}, clip, width=\columnwidth]{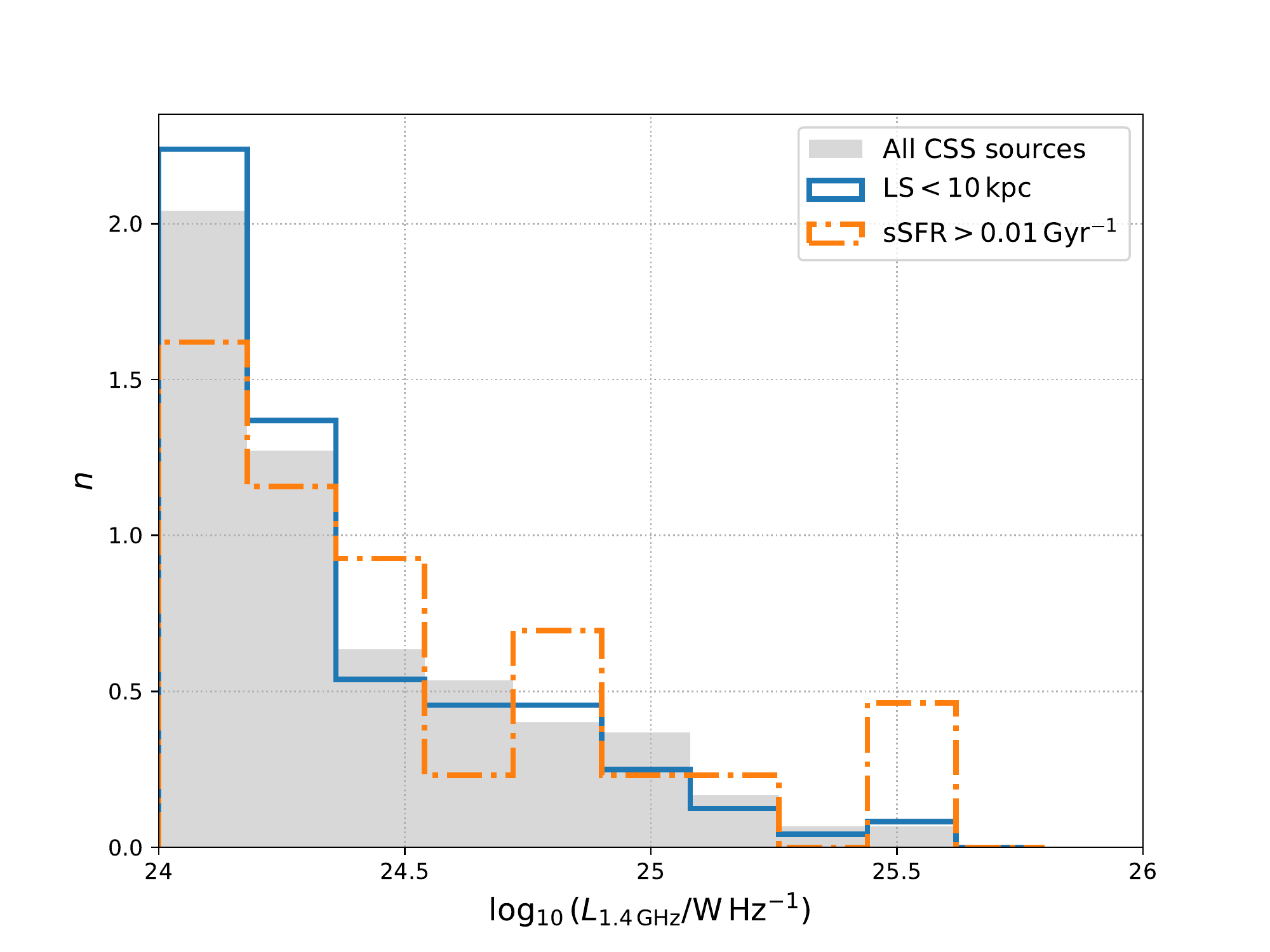}}
    \caption{Normalised distributions of $L_{1.4\,\text{GHz}}$ for subsamples of our CSS sources.
    Both panels show the full CSS sample as a solid grey histogram, and the CSS sources with $\text{LS}<10\,\text{kpc}$ as blue solid line.
    Panel a shows a comparison with sources having $\text{LS}\geq10\,\text{kpc}$ (red dashed line), while Panel b shows the radio luminosities of CSS sources with $\text{sSFR}>0.01\,\text{Gyr}^{-1}$ (orange dot-dashed line).}
    \label{fig:luminosity}
\end{figure}

The increase in radio luminosity shown for our toy model jets as the radio source grows is consistent with our data.
In Figure \ref{fig:luminosity}a we show the distributions of $L_{1.4\,\text{GHz}}$ for small ($\text{LS}<10\,\text{kpc}$) and larger ($\text{LS}\geq10\,\text{kpc}$) CSS sources in our sample.
Performing a KS test returns a $p$-value of $8\times10^{-3}$, showing these distributions to be statistically different.
The smaller CSS sources have a median radio luminosity of $10^{24.26}\,\text{W}\,\text{Hz}^{-1}$, while the larger sources have a median value of $10^{24.49}\,\text{W}\,\text{Hz}^{-1}$.
Such a change in luminosity for a $10^{36}\,$W jet would be expected as it grows from a linear size of $\approx 6\,$kpc to $\approx 12\,$kpc (see Figure \ref{fig:jet_models}a).
The $L_{1.4\,\text{GHz}}\,$ distribution of the CSS sources with $\text{sSFR} > 0.01\,\text{Gyr}^{-1}$ (Figure \ref{fig:luminosity}b) is consistent with the luminosity distribution of the smaller CSS sources, having a KS derived $p$-value of $0.26$.

The SF indicators shown in Figure \ref{fig:size_v_ssfr} are visible for different time periods after SF ends.
H$\alpha$ emission is the result of ionisation of the ISM by massive O-type stars, limiting its visibility to $\approx20\,$Myr after the cessation of SF \citep{Kennicutt1998rev}.
Conversely, $D_{4000}$ is affected by the entire stellar population and evolves slowly following a starburst, taking several hundred Myr for a strong break to develop \citep{Goto2008}.
IR colors resulting from SF evolve on a timescale between the two extremes of H$\alpha$ and $D_{4000}$.
The WISE W3 band traces SF through the polycyclic aromatic hydrocarbons associated with B-type stars that live for $\approx 100\,$Myr following a starburst \citep{Peeters2004, Jarrett2011}.

Assuming a jet age on the order of $\sim10\,$Myr, the absence of low $D_{4000}$ values in CSS sources with \text{$\text{LS}\gtrsim10\,$kpc} suggests that the bulk of SF ceased hundreds of Myr prior to the jet being triggered.
Our results are thus inconsistent with the AGN jet triggering the SF unless jet propagation is frustrated for hundreds of Myr.
On the other hand, the presence of strong H$\alpha$ emission in CSS sources with $\text{LS}\lesssim10\,$kpc is indicative of active SF as recently as $10\,$Myr prior to the jet being triggered.
Future observations that measure the jet (a)symmetry and hotspot proper motions are necessary to test if these sources are indeed frustrated.

A tantalising explanation for our results is that of galaxy mergers--a known trigger for both SF and AGN \citep[e.g.][]{Ellison2013, Pearson2019, Gao2020, Pierce2022, Pierce2023}.
In galaxy mergers SF is episodic and the time required for gas to fall into the central engine means that RLAGN are not expected to be triggered until a few hundred Myr after the first starburst \citep{Tadhunter2005, Peirani2010, Shabala2017}.
A final starburst in the merger sequence that ceases $\approx10\,$Myr prior to the jet being triggered, and has a much smaller burst fraction than the initial starburst several hundred Myr earlier, might produce the observed H$\alpha$ with a limited impact on $D_{4000}$.
It is therefore prudent to ask if our sample of CSS sources with enhanced SF are associated with mergers?
To this end we visually inspect optical images obtained from the 9th data release of the Dark Energy Spectroscopic Instrument Legacy Imaging Surveys \citep[][]{Dey2019} for the $24$ CSS sources with $\text{sSFR}>0.01\,\text{Gyr}^{-1}$.
We find that $11\ (46\,\%)$ show clear evidence of tidal features indicative of a recent major galaxy-galaxy interaction.
This is a higher incidence than the $28\,\%$ of the LERG population shown to have tidal features in \citet{Gordon2019}.
The relatively high fraction of our sample with tidal features suggests that mergers likely explain at least some CSS sources with enhanced SF, and this warrants further study.

\subsection{Are our CSS Sources Really Variable Sources?}

We have selected our CSS sources using legacy data from two different radio surveys.
Because these observations were not made simultaneously--two decades separate FIRST and VLASS--it is possible the difference in flux density measurements may be an effect of radio source variability rather than the shape of spectral energy distribution.
Variable radio sources typically have very compact morphologies. 
\citet{Nyland2020} show that sources showing variability between FIRST and VLASS have $\text{LS}<1\,$kpc, while \citet{Wolowska2021} use VLBI imaging to show such sources typically have sizes of just a few tens of parsecs.

Of the $24$ CSS sources with $\text{sSFR} > 0.01\,\text{Gyr}^{-1}$, only $3$ are completely unresolved by VLASS (shown as upper limits in Figure \ref{fig:size_v_ssfr}).
A further $9$ of these $24$ sources have measured linear sizes below $2\,$kpc, notably all of which are greater than $1\,$kpc.
The other half of our CSS sample with enhanced SF have \text{$2\,\text{kpc} < \text{LS} < 8\,\text{kpc}$} and are therefore larger than variable sources are expected to be.
If we were to cautiously assume that the $12$ sources within our sample with enhanced SF and $\text{LS}<2\,$kpc are all variable sources then our conclusion would still be valid: CSS sources with enhanced SF are smaller than $\approx 10\,$kpc.

\section{Conclusions} 
\label{sec:conclusions}


In this Letter we have systematically investigated the relationship between star formation and radio source size in CSS sources.
We find that where enhanced SF is present the radio source has $\text{LS} \lesssim 10\,$kpc, while passive hosts are seen in CSS sources with $0 \leq \text{LS} < 20\,$kpc.
Based on simulated jet propagation times, the absence of CSS hosts with weak $4,000\,$\AA\ break strengths at $\text{LS}\gtrsim10\,$kpc suggests the bulk of SF ceased several hundred Myr before the AGN jet was triggered.
The presence of H$\alpha$ emission in CSS sources with $\text{LS} < 10\,$kpc indicates that some SF occurred $\approx 10\,$ Myr prior to the jet triggering.
We interpret this apparent ambiguity as being the result of episodic SF in these CSS sources where the later starbursts have a lower `burst fraction', potentially resulting from galaxy-galaxy interactions.

\section*{}\noindent 
We thank the anonymous referee for their helpful report that has improved the quality of this work.
Y.A.G. is supported by U.S. National Science Foundation grant AST 20-09441.
C.P.O., S.A.B. and C.D. are supported by NSERC, the National Sciences and Engineering Research Council of Canada.

%

\facilities{
This work made use of observations from the Very Large Array (VLA), SDSS and WISE.\\
\indent The VLA is operated by The National Radio Astronomy Observatory, a facility of the National Science Foundation operated under cooperative agreement by Associated Universities, Inc.\\
\indent Funding for the SDSS and SDSS-II has been provided by the Alfred P. Sloan Foundation, the Participating Institutions, the National Science Foundation, the U.S. Department of Energy, the National Aeronautics and Space Administration, the Japanese Monbukagakusho, the Max Planck Society, and the Higher Education Funding Council for England.
The SDSS Web Site is \url{http://www.sdss.org/}.\\
\indent The SDSS is managed by the Astrophysical Research Consortium for the Participating Institutions. 
The Participating Institutions are the American Museum of Natural History, Astrophysical Institute Potsdam, University of Basel, University of Cambridge, Case Western Reserve University, University of Chicago, Drexel University, Fermilab, the Institute for Advanced Study, the Japan Participation Group, Johns Hopkins University, the Joint Institute for Nuclear Astrophysics, the Kavli Institute for Particle Astrophysics and Cosmology, the Korean Scientist Group, the Chinese Academy of Sciences (LAMOST), Los Alamos National Laboratory, the Max-Planck-Institute for Astronomy (MPIA), the Max-Planck-Institute for Astrophysics (MPA), New Mexico State University, Ohio State University, University of Pittsburgh, University of Portsmouth, Princeton University, the United States Naval Observatory, and the University of Washington.\\
\indent WISE is a joint project of the University of California, Los Angeles, and the Jet Propulsion Laboratory/California Institute of Technology, and NEOWISE, which is a project of the Jet Propulsion Laboratory/California Institute of Technology. WISE and NEOWISE are funded by the National Aeronautics and Space Administration.}


\software{We used the following software packages in the preparation of this manuscript:
AstroPy \citep{Astropy2013, Astropy2018, Astropy2022},
Matplotlib \citep{Hunter2007},
NumPy \citep{Harris2020},
SAOImage DS9 \citep{Joye2003},
SciPy \citep{Virtanen2020},
Seaborn \citep{Waskom2021}
and TOPCAT \citep{Taylor2005}.}

\bibliography{css_sf.bib}{}
\bibliographystyle{aasjournal}

\end{document}